\documentclass{llncs}

\usepackage{amsmath}
\usepackage{amssymb}
\usepackage{latexsym}
\usepackage{graphicx}

\usepackage{mdwmath}
\usepackage{mdwtab}
\usepackage[tight,footnotesize]{subfigure}

\usepackage {bsymb}

\newcommand\bcdot{\ensuremath{
  \mathchoice
   {\mskip\thinmuskip\lower0.2ex\hbox{\scalebox{1.5}{$\cdot$}}\mskip\thinmuskip}}
   {\mskip\thinmuskip\lower0.2ex\hbox{\scalebox{1.5}{$\cdot$}}\mskip\thinmuskip}
   {\lower0.3ex\hbox{\scalebox{1.2}{$\cdot$}}}
   {\lower0.3ex\hbox{\scalebox{1.2}{$\cdot$}}}
}

\begin{document}

\title{Development of Fault Tolerant MAS \\with Cooperative Error Recovery \\by Refinement in Event-B}

\titlerunning{Development of Fault Tolerant MAS with Cooperative Error Recovery}

\authorrunning{Inna Pereverzeva et al.}

\author{Inna Pereverzeva$^{1,2}$ \and Elena Troubitsyna$^{2}$ \and Linas Laibinis$^{2}$}

\authorrunning{Inna Pereverzeva et al.}

\institute{Turku Centre for Computer Science \and \AA{}bo Akademi University, Finland \\ 
\email{\textbraceleft inna.pereverzeva, elena.troubitsyna, linas.laibinis\textbraceright@abo.fi}}

\maketitle

\begin{abstract}
Designing fault tolerance mechanisms for multi-agent systems is a notoriously difficult task.  In this paper we present an approach to formal development of a fault tolerant multi-agent system by refinement in Event-B. We demonstrate how to formally specify cooperative error recovery and dynamic reconfiguration in Event-B. Moreover, we discuss how to express and verify essential properties of a fault tolerant multi-agent system while refining it. The approach is illustrated by a case study -- a multi-robotic system.
\end{abstract}

\section{Introduction}
\label{sec:intro}
\vspace*{-0.2cm}
Multi-agent systems (MAS) and in particular the agent cooperation have been a subject of an active research over the last decade. In this paper we focus on studying the fault tolerance aspects of agent cooperation. Namely, we discuss how to express and verify essential properties of a fault tolerant MAS. Moreover, we show by example how to formally derive a specification of a MAS that relies on dynamic reconfiguration and cooperative error recovery to achieve fault tolerance.

In this paper,  we present a formal development of a cleaning multi-robotic system. 
The system has a heterogenous architecture consisting of several stationary devices, base stations,  that coordinate the work of respective groups of robots.  Since both base stations and robots can fail, the main objective of our formal development is to formally specify cooperative error recovery and verify that the proposed design ensures goal reachability.  The proposed development approach ensures goal reachability ``by construction". It is based on refinement in Event-B~\cite{EventB} -- a formal top-down approach to correct-by-construction system development. In this paper we demonstrate how to formally define a system goal and, in a stepwise manner, \emph{derive} a detailed specification of the system architecture.

The paper is structured as follows. In Section 2 we briefly overview the Event-B formalism. In Section 3 we define the main principles of formal reasoning about goal-oriented MAS, describe the requirements for our case study -- a multi-robotic system -- and outline the development strategy. Section 4 presents a formal development of the system and demonstrates how to express and verify its properties during the refinement process. Finally, in Section 5 we overview the related work and discuss the achieved results.

\vspace*{-0.2cm}
\section{Modelling and Refinement in Event-B}
\label{sec:eventb}
\vspace*{-0.2cm}
Event-B is a state-based formal approach that promotes the \linebreak correct-by-construction development paradigm and formal verification by theorem proving~\cite{EventB}.
In Event-B, a system model is specified using the notion of an \emph{abstract state machine}. An abstract state machine encapsulates the model state represented as a collection of variables, and defines operations on this state, i.e., it describes the \textit{behaviour} of the modelled system. A machine may have the accompanying component, called \emph{context}. A context may include user-defined carrier sets, constants and their properties (model axioms). In Event-B, the model variables are strongly typed by the constraining predicates called \textit{invariants}. Moreover, the invariants specify important properties that should be preserved during the system execution.

The dynamic behaviour of the system is defined by the set of atomic \textit{events}. Generally, an event can be defined as
\vspace*{-0.15cm}
\begin{small}
$$\textbf{evt}\;\widehat{=}\;{\textsf{\bf any}}\ vl\ {\textsf{\bf where}}\ g\ {\textsf{\bf then}}\ S\ {\textsf{\bf end}}
\vspace*{-0.2cm}
$$
\end{small}
\noindent where $vl$ is a list of new local variables, $g$ is the \textit{guard}, and $S$ is the \textit{action}. The guard is a state predicate that defines the conditions under which the action can be executed. In general, the action of an event is a parallel composition of deterministic or non-deterministic assignments.

The Event-B refinement process allows us to gradually introduce implementation details, while preserving functional correctness. The consistency of Event-B models, i.e., invariant preservation, correctness of refinement steps, should be formally demonstrated by discharging relevant proof obligations.
The verification efforts, in particular, automatic generation and proving of the required proof obligations, are significantly facilitated by the Rodin platform~\cite{RODINPLAT}. Proof-based verification as well as reliance on abstraction and decomposition adopted in Event-B offers the designers a scalable support for the development of such complex distributed systems as MAS.

\vspace*{-0.2cm}
\section{Multi-Agent Systems}
\vspace*{-0.2cm}
Our paper focuses on formal modelling and development of MAS that should function autonomously, i.e., without human intervention. 
Typically, the main task or \emph{goal} that such a MAS should accomplish is split between the deployed agents.
Since agents may fail, to ensure success of the overall goal, we should incorporate some fault tolerance mechanisms into the system design. These mechanisms rely on cooperative error recovery that allows the system dynamically reallocate functions from the failed agents to the healthy ones. A large number of failure modes and scenarios makes verification of goal reachability in the presence of cooperative error recovery quite difficult and time-consuming. Therefore, there is a clear need for rigorous approaches that support scalable design and verification in a systematic manner.

\vspace*{-0.2cm}
\subsection{Towards a Formalisation of a Goal-Oriented MAS}
\vspace*{-0.2cm}
Let us now describe more formally the properties that a MAS is expected to satisfy. 
\begin{enumerate}

\item Let us to denote the system state space as $\Sigma$. Then the main goal $G$ that the system aims at accomplishing can be associated with a specific predicate over $\Sigma$:
\vspace*{-0.1cm}\begin{small}
$$\displaystyle G\ :\ \Sigma\ \rightarrow BOOL.\vspace*{-0.5cm}$$
\end{small}

\noindent In other words, the system goal is reached in a particular state $\sigma$ if and only if $G(\sigma) = \textit{TRUE}$.
\item The system goal $G$ can usually be decomposed into a set of subgoals $SG_i,$ where $i \in 1..n$. We suppose that there exists a precise relationships, $\textit{Expr}$, between reachability of the main goal and that of the subgoals such that:
\begin{small}
$$\displaystyle G(\sigma) = \textit{TRUE}\ \Leftrightarrow\ \textit{Expr}(SG_1(\sigma), ..., SG_n(\sigma)) = \textit{TRUE}.\vspace*{-0.2cm}$$
\end{small}
\item We assume that the system is \emph{stable} with respect to its goals (subgoals), i.e., once a particular goal (subgoal) is reached, it stays reached. In B models, this property can be formulated as an invariant (using auxiliary variables to refer to the relevant part of the previous system state $\sigma_{prev}$) of the form:
\begin{small}
$$\displaystyle G(\sigma_{prev}) = \textit{TRUE}\ \Rightarrow\ G(\sigma) = \textit{TRUE}.\vspace*{-0.2cm}$$
\end{small}
\item In multi-agent systems, (sub)goals are usually achieved by system agents. Often a specific (sub)goal should be accomplished only by a particular subset of agents. We call such agents $\textit{eligible}$. Formally, for each subgoal $SG_i$, we define a eligibility function, $SG_i\_Elig$:
\vspace*{-0.1cm}\begin{small}
$$\displaystyle SG_i\_Elig\ :\ \textit{AGENTS}\times \Sigma\ \rightarrow\ \textit{BOOL},\vspace*{-0.6cm}$$
\end{small}

\noindent where $AGENTS$ denotes a set of all the system agents. In practice, such a function often checks whether a particular agent  belongs to a specific class of agents responsible for achieving the subgoal. Moreover, it also determines whether the agent is able to perform the required task, i.e., it has not failed.

\item  Since MAS are distributed, we assume that the knowledge about the (sub)goal
 reachability is shared among the agents.  In other words, each agent has its own local copy of it. We model this by a family of functions $\textit{Agent}\_SG_i$, where $i \in 1..n$:
\vspace*{-0.1cm}\begin{small}
$$\displaystyle \textit{Agent}\_SG_i\ :\ \textit AGENTS \times \Sigma\ \rightarrow\ \textit{BOOL}.\vspace*{-0.4cm}$$
\end{small}

The local and global knowledge must be consistent, i.e.,
\vspace*{-0.05cm}\begin{small}
\begin{equation}
\label{eqno1}
SG_i(\sigma) = \textit{FALSE}\ \Rightarrow\ \forall a\in \textit{AGENTS}.\: \textit{Agent}\_SG_i(a,\sigma) = \textit{FALSE}.
\vspace*{-0.55cm}
\end{equation}
\end{small}

In practice, it means that the information about reaching a particular subgoal by an agent should be broadcasted to the other agents.

\item  The essential property of the considered MAS is eventual reachability of its main goal. In B models, such reachability is typically abstractly modelled by a single event reaching the desired system state. The event is then refined by the group of events terminating in the desired state. To prove termination, the natural number expression, \emph{variant}, should be defined and shown to be decreased by the  refined events.  We assume that there exists a variant expression $V_i, V_i\in \Sigma\rightarrow \textit{NAT}$, for each subgoal $SG_i$ of the system.

Since system agents may fail before reaching the assigned (sub)goal, to prove eventual goal reachability, we need to introduce various agent cooperative recovery scenarios that allow the active agents to take over the failed ones. We will consider several such scenarios  later in this paper.

\end{enumerate}

To exemplify a goal-oriented development of  MAS, next we present  our case study -- a multi-robotic system. We start by informally defining the system requirements. Then we demonstrate how to formally develop such a system in Event-B and prove its essential properties.

\label{sec:CaseS}
\vspace*{-0.2cm}
\subsection{A Case Study: A Multi-Robotic System}
\vspace*{-0.2cm}
The goal of the multi-robotic system is
to get a certain territory cleaned by the robots. The territory is divided into several \textit{zones},
which in turn are further divided into a number of \textit{sectors}. Each zone has a \textit{base station} that
coordinates the cleaning activities within the zone. In general, one base station might coordinate several zones.
In its turn, each base station supervises a number of robots attached to it by assigning cleaning tasks to them.

A robot is an autonomous electro-mechanical device that can move and clean. A base station may assign
a robot a specific sector to clean. Upon receiving the task, the robot autonomously moves to this sector and
performs cleaning. After successfully completing its mission, the robot returns back to the base station
to receive a new task. The base station keeps track of the cleaned and non-cleaned sectors.  Moreover, the base stations periodically exchange the information
about their cleaned sectors.

While performing the given assignment, a robot may fail. Subsequently it leads to a failure to clean the assigned sector. We assume that a base station is able to detect all the failed robots attached to it. In  case of a robot failure, the base station may assign another active robot to perform the failed task.

A base station might fail as well. We assume that a failure of a base station  can be detected by the others stations. In that case, the healthy base stations redistribute control over the zones and robots coordinated by the failed station.

Let us now formulate the main requirements and properties associated with the multi-robotic system that is informally described above.
\begin{description}
 \item[$\mathtt{(PR1)}$] \textit{The main system goal: the whole territory has to be cleaned.}
 \item[$\mathtt{(PR2)}$] \textit{To clean the territory, every its zone has to be cleaned.}
 \item[$\mathtt{(PR3)}$] \textit{To clean a zone, every its sector has to be cleaned.}
 \item[$\mathtt{(PR4)}$] \textit{Every cleaned sector (zone) remains cleaned during the system execution.}
 \item[$\mathtt{(PR5)}$] \textit{No two robots should clean the same sector.}
In other words, a robot gets only non-assigned and non-cleaned sectors to clean.
 \item[$\mathtt{(PR6)}$] \textit{The information about the cleaned sectors stored in any base station has to be consistent with the current state of the territory.}
More specifically, if a base station considers a particular sector in some zone to be cleaned, then  this sector is marked as cleaned in the memory of the base station responsible for it. Also, if a sector is marked as non-cleaned in the memory of the base station responsible for it, then any base station sees it as non-cleaned.
 \item[$\mathtt{(PR7)}$] \textit{Base station cooperation: if a base station has been detected as failed then some base station will
take the responsibility for all the zones and robots of the failed base station. }
 \item[$\mathtt{(PR8)}$] \textit{Base station cooperation: if a base station has no more active robots, a group of robot is sent to this base station from another base station.}
 \item[$\mathtt{(PR9)}$] \textit{Base station cooperation: if a base station has cleaned all its zones, its active robots may be reallocated under control of  another base station.}
\end{description}

The last three requirements essentially describe the cooperative recovery mechanisms that we assume to be present in the described multi-robot system.

\vspace*{-0.2cm}
\subsection{Formal Development Strategy}
\vspace*{-0.2cm}
In the next section we will present a formal Event-B development of the described multi-system robotic system. We demonstrate how to specify and verify the given properties (PR1)--(PR9). Let us now give a short overview of this development and highlight formal techniques used to ensure the proposed properties.

We start with a very abstract model, essentially representing the system behaviour as a process iteratively trying to achieve the main goal (PR1). The next couple of data refinement steps decompose the main goal into a set of subgoals, i.e., reformulate it in terms of zones and sectors. We will define the gluing invariants establishing a formal relationship between goals and the corresponding subgoals. Thus, we will define a relation $Expr$, described in Section 3.1.

While the specification remains highly abstract,  we postpone the proof of goal reachability property by defining  the corresponding events as \textit{anticipated}. Once, as a result of the refinement process, the model becomes sufficiently detailed, we change the event statuses into \textit{convergent} and prove their termination. To achieve this, we need to define a \textit{variant} -- a natural number expression -- and show that the execution of any of these events decreases it. 

Next we introduce the agent types -- base stations and robots. The base stations coordinate execution of the tasks required to achieve the corresponding subgoal, while the robots execute the tasks allocated to them. We formally define the relationships between different types of agents, as well as agents and respective subgoals. These relationships are specified  and proved as invariant properties of the model.

The consequent refinement steps explicitly introduce agent failures, the information exchange as well as cooperation activities between the agents. The integrity between the local and the global information stored within base stations is again formulated and proved as model invariant properties.

We assume that communication between the base stations as well as the robots and the base stations is reliable. In other words, messages are always transmitted correctly without  any loss or errors. The main focus of our development is on specifying and verifying the cooperative recovery mechanisms.

\vspace*{-0.2cm}
\section{Development of a Multi-Robotic System in Event-B}
\vspace*{-0.2cm}
\subsection{Modelling system goals and subgoals}
\vspace*{-0.1cm}
\paragraph{\bf Abstract model.}Our initial model abstractly represents the behaviour of the described multi-robotic system.
We aim at ensuring the property $\mathtt{(PR1)}$. We define a variable $goal \in STATE$ that models the current state of the system goal,
where $STATE\!=\!\{incompl,$ $compl\}$. In the process of achieving the goal, modeled by the event $\mathsf{Body}$, the variable $goal$ may eventually change its value from $incompl$ to $compl$. The value $compl$ corresponds to the situation when the goal is achieved, i.e., the whole territory is cleaned.
 The system continues its execution until the whole territory is not cleaned, i.e., while $goal$ stays $incompl$.
 \vspace*{-0.4cm}
 \begin{center}
 \begin{scriptsize}
 \fbox{
 \parbox[center]{6.5cm}{
$\mathsf{Body}\;\widehat{=}$
$\textbf{status} \; anticipated $  \\
 \hspace*{5pt} $\textbf{when}\ goal \neq compl\ \textbf{then}\ goal :\in STATE \ \textbf{end}$\\
 \vspace*{-0.2cm}
 }}
 \end{scriptsize}
 \end{center}

\vspace*{-0.3cm}
\paragraph{\bf First refinement.} In our first refinement step we elaborate on the process of cleaning the territory. Specifically, we assume that the whole territory is divided into $n$ zones, where $n \in \mathbb{N}$ and $n \geq 1$, and aim at ensuring  the property $\mathtt{(PR2)}$.
We augment our model with a representation of subgoals. We associate the notion of a \textit{subgoal} with the
process of \textit{cleaning a particular zone}. A subgoal is achieved only when the corresponding zone is cleaned.
A new variable $zones$ represents the current subgoal status for every zone:
$\displaystyle zones \in 1..n \tfun STATE.$

To establish the relationship between goal and subgoals and  formalise the property $\mathtt{(PR2)}$ per se, we formulate the gluing invariant:
\vspace*{-0.15cm}\begin{small}
$$\displaystyle goal=compl \leqv  zones[1..n]=\{compl\}.\vspace*{-0.15cm}$$
\end{small}
The invariant can be understood as follows: the territory is considered to be cleaned if and only if its every zone is cleaned. In this case, the $Expr$, defined in the Section 3, becomes a conjunction of the subgoals.
To model cleaning of a zone(s), we refined the abstract event $\mathsf{Body}$. We model it in such a way that, while a certain subgoal is reached, it stays reached. Hence we ensure the property $\mathtt{(PR4)}$.
\vspace*{-0.25cm}
\paragraph{\bf Second refinement.} Next we further decompose system subgoals into a set of subsubgoals.
We assume that each zone in our system is divided into $k$ sectors, where $k \in \mathbb{N}$ and $k \geq 1 $, and aim at formalising the property $\mathtt{(PR3)}$.
We establish the relationship between the notion of a subsubgoal (or simply \textit{a task}) and the process of
\textit{cleaning a particular sector}. A task is completed when the corresponding sector is cleaned.
A new variable $territory$ represents the current status of each sector:
\vspace*{-0.2cm}\begin{small}
$$\displaystyle territory \in  1\upto n \tfun  (1\upto k \tfun  STATE).\vspace*{-0.55cm}$$
\end{small}

\noindent The following gluing invariant expresses the relationship between subgoals and subsubgoals (tasks) and correspondingly
ensures the property $\mathtt{(PR3)}$:
\vspace*{-0.1cm}\begin{small}
$$\displaystyle
\forall  j\qdot  j \in  1\upto n \limp  (zones(j)=compl \leqv  territory(j)[1\upto k]=\{ compl\}).\vspace*{-0.15cm}
$$
\end{small}
The invariant says that a zone is cleaned if and only if each of its sectors is cleaned.

The refined event $\mathsf{Body}$ is now models cleaning of a previously non-cleaned
sector:
\vspace*{-0.4cm}
\begin{center}
\begin{scriptsize}
\fbox{
\parbox[center]{9cm}{
$\textbf{Body}$ $\;\widehat{=}$ refines $\textbf{Body}$ status $\textbf{anticipated}$\\
\hspace*{1pt} $\textbf{any} \; \; z, s, result$  \\
\hspace*{1pt} $\textbf{when}  \; \; z \in 1..n  \wedge s \in 1\upto k \wedge territory(z)(s)\neq compl \ \wedge result \in  STATE$  \\
\hspace*{1pt} $\textbf{then} \; \; \;territory(z) :=  territory(z) \ovl  \{ s\mapsto result\} \;\; \textbf{end}$
}
}
\end{scriptsize}
\end{center}
Let us observe that the event $\mathsf{Body}$ also preserves the property $\mathtt{(PR4)}$.

\vspace*{-0.2cm}
\subsection{Introducing Agents}
\vspace*{-0.1cm}
In the third refinement step  we augment our model with a representation of agents. In the model context, we define the abstract finite set $AGENTS$ and its disjointed non-empty subsets
$RB$ and $BS$ that represent the robots and the base stations respectively. To define a relationship between
a zone and its supervising base station, we introduce the variable $responsible$:
\vspace*{-0.1cm}
\begin{small}
$$\displaystyle responsible \in  1\upto n \tfun  BS.\vspace*{-0.55cm}$$
\end{small}

Each active robot is supervised by a certain base station.
We model this relationship between robots and
their supervised station by a variable $attached$, defined as a partial function:
\vspace*{-0.2cm}\begin{small}
$$\displaystyle attached \in  RB \pfun  BS.\vspace*{-0.55cm}$$
\end{small}

To coordinate the cleaning process, a base station stores the information about its own cleaned sectors and updates
the information about the  status of the other cleaned sectors. We assume that each base station has a ``map" -- the knowledge about
all sectors of the whole territory.  To model this, we introduce a new variable $local\_map$:
\vspace*{-0.1cm}\begin{small}
$$\displaystyle
local\_map \in  BS \tfun  (1\upto n \pfun   (1\upto k \tfun  STATE)).\vspace*{-0.55cm}$$
\end{small}

The abstract variable $territory$ represents the global knowledge on the whole territory.
For any sector and zone, this global knowledge  has to be consistent with the information stored by the base stations.
In particular, if in the local knowledge of any base station a sector is marked as cleaned, then it should be cleaned according to the global knowledge as well.
To establish those relationships, we formulate and prove the following invariant:

\vspace*{-0.6cm}\begin{small}
\begin{multline*}
\forall  bs, z, s \qdot  bs \in  ran(responsible) \land  z \in  1\upto n \land  s \in  1\upto k \ \limp  \\ (territory(z)(s)=incompl \limp  local\_map(bs)(z)(s)=incompl).\vspace*{-0.3cm} \end{multline*}
\end{small}
\vspace*{-0.4cm}

For each base station, the local information about its zones and sectors always coincides with the global knowledge about the corresponding zones and sectors:
\vspace*{-0.4cm}\begin{small}
\begin{multline*} \forall  bs, z, s \qdot  bs \in  ran(responsible) \land  z \in  1\upto n \land  responsible(z)=bs \land  s \in  1\upto k \ \limp \\ (territory(z)(s)=incompl \leqv  local\_map(bs)(z)(s)=incompl).
\end{multline*}\vspace*{-0.5cm}
\end{small}

\noindent All together, these three invariants formalise the property $\mathtt{(PR6)}$. It easy to see that these invariants are special cases of  the property (\ref{eqno1}),
formulated in the Section 3.

A base station assigns a cleaning task to its attached robots. Here, we have to ensure the property $\mathtt{(PR5)}$ -- \textit{no two robots can clean the certain sector at the same time}.  We introduce a number of new variables and an event $\mathsf{NewTask}$ to model this behaviour.

The robot failures have some impact on execution of the cleaning process. The task cannot be performed
if the robot assigned for it has failed. To reflect this behaviour, we refine the event $\mathsf{Body}$
by two events $\mathsf{TaskSuccess}$ and $\mathsf{TaskFailure}$, which respectively model successful and
unsuccessful execution of the task.
\vspace*{-0.4cm}
\begin{center}
\begin{scriptsize}
\fbox{
\parbox[center]{11.9cm}{
$\textbf{TaskSuccess}\;\widehat{=}$ refines $\textbf{Body}$ status $\textbf{convergent}$\\
\hspace*{1pt} $\textbf{any} \; \; bs, rb, z, s$  \\
\hspace*{1pt} $\textbf{when} \; \; bs \in  BS \wedge rb \in  dom(attached) \wedge attached(rb)=bs \ \wedge z \in  1\upto n \ \wedge responsible(z)=bs \wedge $  \\
\hspace*{25pt} $asgn\_z(rb)=z  \wedge s \in  1\upto k \wedge asgn\_s(rb)=s \wedge local\_map(bs)(z)(s)=incompl$  \\
\hspace*{1pt} $\textbf{then} \; \;\;territory(z) :=  territory(z) \ovl  \{ s\mapsto compl\} \parallel$ \\
\hspace*{25pt} $asgn\_s(rb) :=  0 \parallel asgn\_z(rb) :=  0 \parallel counter :=  counter-1 \parallel$ \\
\hspace*{25pt} $local\_map(bs) :=   local\_map(bs) \ovl  \{ z\mapsto local\_map(bs)(z)\ovl \{ s\mapsto compl\} \} $ \; $\textbf{end}$ \\
\vspace*{-0.3cm}
}}
\end{scriptsize}
\end{center}
At this refinement step, we are ready to demonstrate that the events \linebreak $\mathsf{TaskSuccess}$ and $\mathsf{TaskFailure}$ converge.
To prove it, we define the variant expression over system variables, $counter\ +\ card(dom(attached))$, and prove that it is decreased by new events.  An auxiliary variable $counter$ stores the number of all non-cleaned sectors of the whole territory, see~\cite{TUCSTR1052} for details.

A base station keeps track of the cleaned and non-cleaned sectors and repeatedly receives
the information from the other base stations about their cleaned sectors. The knowledge is
inaccurate for the period when the information is sent but not yet received. In this refinement step, we
abstractly model receiving the information by a base station.
We introduce a new event $\mathsf{UpdateMap}$ to model updating of the local map of a base station.

In this refinement step we also introduce an abstract representation of the base station cooperation defined by the property $\mathtt{(PR7)}$.
Namely, we allow to reassign a group of robots from one base station to another.  We define such a behaviour in the event $\mathsf{ReassignRB}$.
In the next refinement steps we will elaborate on this event and define the conditions under which this behaviour takes place.

Additionally, we model a possible redistribution between the base stations their pre-assigned responsibility for zones and robots. This behaviour
 is defined in the new  event $\mathsf{GetAdditionalResponsibility}$ presented below. The guard of the  event defines the conditions when such a change is allowed. A base station can take the responsibility for a set of new zones if it has the accurate knowledge about these zones, i.e., the information about their cleaned and non-cleaned sectors. 
\vspace*{-0.2cm}
\begin{center}
\begin{scriptsize}
\fbox{
\parbox[center]{11.8cm}{
\textbf{$\textbf{GetAdditionalResponsibility}\;\widehat{=}$} \\
\hspace*{1pt} $\textbf{any} \;\; bs\_i, bs\_j, rbs, zs$  \\
\hspace*{1pt} $\textbf{when}  \; \; bs\_i \in  BS \wedge bs\_j \in  BS \wedge zs \subset  1\upto n \ \wedge zs=dom(responsible\ranres \{ bs\_i\} ) \ \wedge bs\_i\neq bs\_j \ \wedge $  \\
\hspace*{24pt} $rbs \subset  dom(attached) \ \wedge rbs=dom(attached\ranres \{ bs\_i\} )  \ \wedge  bs\_j \in  ran(responsible)  \ \wedge  $  \\
\hspace*{24pt} $(\forall  z,s\qdot  z\in zs \land  s\in 1\upto k \limp  territory(z)(s)=local\_map(bs\_j)(z)(s))$  \\
\hspace*{0pt}  $\textbf{then}  \;\;\; responsible :=  responsible \ovl  (zs \cprod  \{ bs\_j\} ) \parallel  attached :=  attached \ovl  (rbs \cprod  \{ bs\_j\} ) \parallel $ \\
\hspace*{24pt} $local\_map(bs\_i) :=  \emptyset  \parallel asgn\_z :=  asgn\_z \ovl  (rbs \cprod  \{ 0\} ) \parallel asgn\_s :=  asgn\_s \ovl  (rbs \cprod  \{ 0\} )$ \\
\hspace*{0pt} $\textbf{end}$ \\
\vspace*{-0.3cm}
}}
\end{scriptsize}
\end{center}
\vspace*{-0.2cm}
Modelling this behaviour allows us to formalise the property $\mathtt{(PR9)}$.

\vspace*{-0.25cm}
\subsection{Modelling of Broadcasting}
\vspace*{-0.1cm}
In next, fourth refinement step we aim at defining an abstract model of broadcasting. After receiving a notification from
a robot about successful cleaning the assigned sector, a base station updates its local map and broadcasts the message about the cleaned sector
to the other base stations. In its turn, upon receiving the message, each base station correspondingly updates its own local map. A new relational variable 
$msg$ models the message broadcasting buffer:
\begin{small}
$$\displaystyle
msg \in  BS \rel  (1\upto n \cprod  1\upto k).\vspace*{-0.cm}$$
\end{small}

\vspace*{-0.5cm}\noindent If a message $(bs\mapsto (z\mapsto s))$ belongs to this buffer then the sector $s$ from the zone $z$ has been cleaned.
The first element of the message, $bs$, determines to which base station  the message is sent.
\noindent If there are no messages in the $msg$ buffer for any particular base station then the
local map of this base station is accurate, i.e., it coincides with the global knowledge about the territory:
\vspace*{-0.2cm}\begin{small}
\begin{multline*}
\forall  bs, z, s \qdot  z \in  1\upto n \land  s \in  1\upto k \land  bs \in  ran(responsible) \land  (bs\mapsto (z\mapsto s)) \notin  msg\limp \\
 territory(z)(s)=local\_map(bs)(z)(s),
\end{multline*}
\vspace*{-0.85cm}
\begin{multline*}
\forall  bs\qdot  bs \in  ran(responsible) \land  bs\notin dom(msg) \limp  \\ (\forall  z,s\qdot  z\in 1\upto n \land  s\in 1\upto k \ \limp territory(z)(s)=local\_map(bs)(z)(s)).\\
\end{multline*}
\end{small}

\vspace*{-0.9cm} After receiving a notification about successful cleaning of a sector, a base station marks this sector as
cleaned in its local map and then broadcasts the message about it to other base stations. To model this, we refine the abstract events $\mathsf{TaskSuccess}$ and $\mathsf{UpdateMap}$.

\vspace*{-0.2cm}
\subsection{Introducing Robot and Base Station Failures}
\vspace*{-0.1cm}
\paragraph{\bf Fifth refinement.} Now we aim at modelling possible robot failures. We elaborate on the abstract events concerning
robot and zone reassigning. We start by partitioning the robots into active and failed ones. The current set of
all active robots is defined by a new variable $active$.
Initially all robots are active, i.e., $active=RB$. A new event $\mathsf{RobotFailure}$ models possible robot failures that can happen
at any time during system execution.
We make an assumption that the last active robot can not fail and add the corresponding guard $card(active) > 1$ to the event $\mathsf{RobotFailure}$
to restrict possible robot failures. In practice, the constraint to have at least one operational agent associated with our model can be validated by probabilistic modelling of goal reachability, which is planned as a future work. Let us also note that for multi-robotic systems with many homogeneous agents this constraint is usually satisfied.

A base station monitors all its robots and detects the failed ones.
The abstract event $\mathsf{TaskFailure}$ abstractly models such robot detection.

To formalise the property $\mathtt{(PR8)}$, we should model a situation when some base station does
not have active robots anymore. In that case, some group of active robots has to be sent to this
base station from another base station. This behaviour is modelled by the event
$\mathsf{ReassignNewBStoRBs}$ that refines the abstract event $\mathsf{ReassignRB}$:
\vspace*{-0.1cm}
\begin{center}
\begin{scriptsize}
\fbox{
\parbox[center]{11.5cm}{
\textbf{$\mathsf{ReassignNewBStoRBs}\;\widehat{=}$} refines $\mathsf{ReassignRB}$\\
\hspace*{1pt} $\textbf{any} \;\; bs\_i, bs\_j, rbs$  \\
\hspace*{1pt} $\textbf{when} \;\; bs\_i \in  BS \wedge bs\_j \in  BS \wedge rbs \subset  active\ \wedge ran(rbs\domres attached)=\{ bs\} \ \wedge rbs \neq  \emptyset \ \wedge  $  \\
\hspace*{24pt} $ran(rbs\domres asgn\_s)=\{ 0\}   \ \wedge  bs\_i \in  ran(responsible) \ \wedge bs\_j \in  ran(responsible) \ \wedge  $  \\
\hspace*{24pt} $bs\_i\neq bs\_j \ \wedge bs\_i \in  ran(rbs\domsub attached) \ \wedge dom(attached\ranres \{ bs\_j\} ) \nsubseteq  active$  \\
\hspace*{1pt} $\textbf{then} \; \; \;attached :=  attached \ovl  (rbs \cprod  \{ bs\_j\} ) \; \; \; \textbf{end}$ \\
\vspace*{-0.3cm}
}}
\end{scriptsize}
\end{center}
This event can be further refined by a concrete procedure to choose a particular base station that will share its robots (e.g., based on load balancing).

Moreover, to ensure the property $\mathtt{(PR9)}$, we consider the situation when all the sectors for which a base station is responsible are cleaned.  In that case, all the active robots of the base station may be sent to some other base station that still has some unfinished cleaning to co-ordinate.  This functionality is
specified by the event $\mathsf{SendRobotsToBS}$ (a refinement of the event $\mathsf{ReassignRB}$).

\vspace*{-0.2cm}
\paragraph{\bf Sixth refinement.}  In the final refinement step presented in the paper, we aim at specifying the base station failures. Each base station might be either
operating or failed. We introduce a new variable $operating \subseteq BS$ to define the set of all operating base stations.
We also introduce a new event  $\mathsf{BaseStationFailure}$ to model a possible base station failure. We again make an assumption that the last active base station can not fail.

In the fourth refinement step we modelled by the event \linebreak $\mathsf{GetAdditionalResponsibility}$ that a base station
can take over the responsibility for the robots and zones of another base station.
Now we can refine this event by introducing an additional condition --  only if a base
station is detected as failed, another base station can take over its responsibility for the respective zones and robots:
\vspace*{-0.2cm}
\begin{center}
\begin{scriptsize}
\fbox{
\parbox[center]{11.8cm}{
\textbf{$\mathsf{GetAdditionalResponsibility}\;\widehat{=}$} refines $\mathsf{GetAdditionalResponsibility}$\\
\hspace*{1pt} $\textbf{any} \; bs\_i, bs\_j, za, rbs$  \\
\hspace*{1pt} $\textbf{when} \; \;bs\_i \in  BS \wedge bs\_j \in  operating \wedge zs \subset  1\upto n \ \wedge zs=dom(responsible\ranres \{ bs\_i\} ) \ \wedge bs\_i\neq bs\_j \ \wedge $  \\
\hspace*{23pt} $rbs \subset  active \ \wedge rbs=dom(attached\ranres \{ bs\_i\} )  \ \wedge bs\_j\notin dom(msg)  \ \wedge bs\_i \notin  operating$  \\
\hspace*{1pt} $\textbf{then} \; \; \;responsible :=  responsible \ovl  (zs \cprod  \{ bs\_j\} ) \parallel attached :=  attached \ovl  (rbs \cprod  \{ bs\_j\} ) \parallel$ \\
\hspace*{23pt} $asgn\_s :=  asgn\_s \ovl  (rbs \cprod  \{ 0\} ) \parallel asgn\_z :=  asgn\_z \ovl  (rbs \cprod  \{ 0\} ) \parallel local\_map(bs\_i) :=  \emptyset $ \\
\hspace*{1pt} $\textbf{end}$ \\
\vspace*{-0.3cm}
}}
\end{scriptsize}
\end{center}
\vspace*{-0.1cm}
As a result of the presented refinement chain, we arrived at a centralised model of the multi-robotic system. We can further refine the system to derive its distributed implementation, relying on the modularisation extension of Event-B to achieve this.

To verify correctness of the models we discharged more than 230 proof obligations. Around 80\% of them have been proved automatically by the Rodin platform and the rest
have been proved manually in the Rodin interactive proving environment.

\vspace*{-0.3cm}
\enlargethispage{0.7cm}
\section{Conclusions and Related Work}
\label{sec:concl}
\vspace*{-0.2cm}
Formal modelling of MAS has been undertaken in~\cite{Roman1,Roman2}. The authors
have proposed an extension of the Unity framework to explicitly define such concepts as
mobility and context-awareness. Our modelling have pursued a different goal -- we have aimed at
formally guaranteeing that the specified agent behaviour achieves the pre-defined
goals.

Formal modelling of fault tolerant MAS in Event-B has been also undertaken by Ball and
Butler~\cite{BaBu09}. They have proposed a number of informally described patterns that
allow the designers to incorporate well-known (static) fault tolerance mechanisms into formal models.
In our approach, we have implemented a more advanced fault tolerance scheme that relies on goal
reallocation  and dynamic reconfiguration to guarantee goal reachability.

The foundational work on goal-oriented development has been done by van
Lamsweerde~\cite{Lam01}. The original motivation behind the goal-oriented development was
to structure the system requirements and derive properties in the form of temporal logic
formulas. Over the last decade, the goal-oriented
approach has received several extensions that allow the designers to link it
with formal modelling~\cite{GOE-B2,GOE-B3,GOE-B1}. These works aimed at
expressing temporal logic properties in Event-B. In our work, we have relied on goals  to
facilitate structuring of the system behaviour and derived a detailed system model that
satisfies the desired properties by refinement.

The theoretical aspects of modelling reachability has been
studied in~\cite{ABR11}. A work similar to our but in the context of discovering a
distributed topology is presented in ~\cite{ABR09}. In our work, reasoning about
liveness property has been put in the context of goal-oriented development.

In this paper we have presented an approach to formal development of a fault tolerant MAS with cooperative error recovery
by refinement in Event-B. The formal development has allowed us to uncover missing
requirements and rigorously define the relationships between agents. It has also facilitated a systematic derivation of 
a complex mechanism for cooperative error recovery.

Our approach has demonstrated a number of advantages comparing to various process-algebraic approaches used
for modelling MAS. We relied on a proof-based verification that allowed us to derive a quite complex
model of the behaviour of a multi-agent robotic system. We did not need to impose restrictions on the size of the model,
number of agents etc. We could comfortably express intricate relationships between the system goals and the employed agents.
Therefore, we believe that Event-B and the associated tool set will provide a suitable framework for formal modelling of complex MAS.

\vspace*{-0.15cm}\bibliographystyle{splncs03}
\vspace*{-0.2cm}
\bibliography{refs}
\vspace*{-0.3cm}

\end{document}